\documentclass[aps,prl,reprint,groupedaddress]{revtex4-1}
\usepackage{braket}
\usepackage[]{graphicx}
\usepackage{subfigure}
\usepackage[table]{xcolor}
\usepackage{MnSymbol}
\begin{document}
\title{$SU(N)$ Representation of Mode Dispersion in a Multimode Optical Fiber: \\ Determining Principal Modes for Mode Division Multiplexing}


\author{ Daniel A. Nolan$^{1,4}$, Giovanni Milione$^{2,3,4}$, and Robert R. Alfano$^{2,3,4}$}
\email[]{NolanDA@corning.com}
\affiliation{$^1$Corning, Inc. Sullivan Park, Corning, NY 14830 USA
\\$^2$Institute for Ultrafast Spectroscopy and Lasers, Physics Department, 
\\City College of New York of the City University of New York,160 Convent Ave., New York, NY 10031 USA
\\$^3$Graduate Center of New York of the City University of New York, 365 Fifth Ave., New York, NY 10016 USA
\\$^4$New York State Center for Complex Light, 160 Convent Ave., New York, NY 10031 USA}


\date{\today}

\begin{abstract}
A method is proposed to experimentally determine principal modes for mode division multiplexing in a multimode optical fiber, i.e., increasing optical fiber information capacity via higher-order transverse eigenmodes. Principal modes are a special linear combination of eigenmodes that do not exhibit mode dispersion up to first order in frequency. This method is based on an $SU(N)$ representation of mode dispersion whereby the generators of the corresponding $SU(N)$ Lie algebra, i.e., generalized Gell-Mann matrices, predict higher-order Stokes parameters which can be used to measure principal modes at the optical fiber input and output. Applications of the $SU(N)$ representation to decoherence free subspaces for single photon entanglement in higher-dimensional Hilbert spaces is discussed.
\end{abstract}
 

\maketitle

\label{sec:intro}  

Optical fibers are the backbone of the modern and future communication infrastructure for both classical and quantum information \cite{kahn2001bottleneck, PhysRevLett.81.5932}. Continuous and exponentially growing demand on optical fiber information capacity is exhausting state of the art methods to maintain information capacity levels, the most prominent being the use of light's wavelength degree of freedom whereby each wavelength serves as an additional information channel, i.e., wavelength division multiplexing (WDM). Corresponding Shannon information capacity limits are directly linked to a conflict between WDM and inherent optical fiber non-linearities \cite{mitra2001nonlinear}. Therefore, the prevention of an impending information ``bottleneck"  will require new methods to increase information capacity. One potential method is the exploitation of light's unused spatial degree of freedom via higher-order transverse modes which span a complete, orthogonal, and high-dimensional Hilbert space \cite{ PhysRevLett.88.013601,gibson2004free}. Using each optical fiber eigenmode as an additional information channel is referred to as mode division multiplexing (MDM) and can potentially increase information capacity in an amount proportional to the number of modes used \cite{richardson2013space}. 

One of the greatest problems in MDM is mode dispersion \cite{arik2013effect}. Mode dispersion is the extension of \textit{polarization mode dispersion} in a single mode optical fiber to the higher-order eigenmodes of a multimode optical fiber. In MDM a light signal launched at the optical fiber input as an eigenmode will exhibit random power transfer back and forth between eigenmodes, i.e., mode coupling, due to imperfections in fabrication and environmental perturbation such as twisting, bending, and changing temperature. As dictated by Maxwell's vector wave equation each eigenmode has a different propagation constant \cite{snyder1983optical}. This, in concert with mode coupling, causes group velocity dispersion at the optical fiber output and in turn detrimental signal errors. 

Remarkably, there exist certain linear combinations of eigenmodes, referred to as \textit{principal modes}, that when launched at the optical fiber input  do not exhibit mode dispersion up to first order in frequency $\omega$ \cite{fan2005principal}. Principal modes are a direct extension of \textit{principal states of polarization} in a single mode optical fiber to the higher-order eigenmodes of a multimode optical fiber. Ideally, using principal modes in place of eigenmodes for MDM can circumvent the problem of mode dispersion. The challenge lies in determining the correct linear combination of eigenmodes to experimentally generated and detect principal modes.
 
In this work, a method is proposed to experimentally determine principal modes for MDM. This method is based on an $SU(N)$ representation of mode dispersion whereby the generators of the corresponding $SU(N)$ Lie algebra, i.e., generalized Gell-Mann matrices, predict higher-order Stokes parameters which can be used to measure principal modes at the optical fiber input and output. Applications of the $SU(N)$ representation to decoherence free subspaces for single photon entanglement in higher-dimensional Hilbert spaces is discussed.

A multimode optical fiber supporting the propagation of N eigenmodes can be represented by the $\hat U$ an $N \times N$ unitary Jones matrix ($\hat U \hat U^{\dag} = \hat I$). A single eigenmode or linear combination thereof at the optical fiber input can be represented by $\ket{s}$ an $N \times 1$ Jones vector. The transformation of $\ket{s}$ by $\hat U$ is given by $\ket{t} = \exp(\imath \phi_o) \hat U \ket{s}$ where $\phi_o$ is a common phase and $\ket{t}$ is an $N \times 1$ vector representing a different linear combination of eigenmodes at the optical fiber output. There exist $\ket{m_s}$, i.e., a special linear combination of optical fiber eigenmodes, that when launched at the optical fiber input do not exhibit mode dispersion up to first order in $\omega$. $\ket{m_s}$ are referred to as \textit{input} principal modes and are eigenstates of the Hermitian operator $ -\imath \hat U^{\dagger} \hat U_{\omega}$ $(\hat{U}_{\omega} = \partial_{\omega} \hat U)$ referred to as the mode dispersion operator, i.e., $ -\imath \hat U^{\dagger} \hat U_{\omega} \ket{m_s} = (\tau +\tau_o) \ket{m_s}$ \cite{fan2005principal}. The transformation of $\ket{m_s}$ by $\hat U$ is given by $\ket{m_t} = \exp(\imath \phi_o) \hat U \ket{m_s}$; $\ket{m_t}$ are referred to as \textit{output} principal modes. MDM with principal modes would involve launching a light signal as $\ket{m_s}$ at the optical fiber input and detecting $\ket{m_t}$ at the output. The problem lies in determining the correct linear combination of optical fiber eigenmodes with which to experimentally generate and detect $\ket{m_s}$ and $\ket{m_t}$, respectively, which in turn requires experimentally determining $ -\imath \hat U^{\dagger} \hat U_ {\omega}$ and $\hat U$. 

Polarization mode dispersion and the experimental determination of principal states of polarization in a single mode optical fiber is well understood \cite{Gordon25042000}. In this case $N=2$ and $-\imath U^{\dagger} \hat U_{\omega}$ is a $2\times2$ Hermitian matrix. By merely exploiting its dimensionality and Hermiticity $-\imath U^{\dagger} \hat U_{\omega}$ can be mathematically represented by the group $SU(2)$ and therefore expanded by generators of the corresponding Lie algebra, e.g. the three Paul matrices, given by $-\imath \hat U^{\dagger} \hat U_{\omega} = \sum_{i=1} ^ 3 \tau_i \hat \sigma_i$ where $\tau_i$ are the coefficients of the expansion and $\vec{\sigma} = \{ \hat \sigma_1, \hat \sigma_2, \hat \sigma_3 \}$:
\begin{equation}
\mathrm{ \bf \hat \sigma}_1 = \left(\begin{array}{cc}1 & 0\\ 0 & -1\end{array}\right), \hspace{3mm}
\mathrm{ \bf \hat \sigma}_2 = \left(\begin{array}{cc}0 & 1 \\1 & 0\end{array}\right), \hspace{3mm}
\mathrm{ \bf \hat \sigma}_3 = \left(\begin{array}{cc}0 & -\imath \\ \imath & 0\end{array}\right).
\end{equation}
The benefit of the $SU(2)$ representation of $- \imath \hat U^{\dagger} \hat U_{\omega}$ is in its ability to connect the Pauli matrices of Eq. 1 to Stokes parameters. As $\ket{m_s}$ is an eigenstate of $-\imath \hat U^{\dagger} \hat U_{\omega}$ its eigenvalue equation is expressed in terms of the expansion by $( \vec \tau_s \cdot \hat \sigma) \ket{m_s} = (\tau +\tau_o) \ket{m_s}$ where $\vec \tau_s$ is a Stokes vector  given by $\vec{ \tau_s } = \tau \braket{ m_s | \vec{ \sigma } | m_s} = \{ \tau_1, \tau_2, \tau_3 \} $ referred to as the \textit{polarization mode dispersion vector}; the components of $\vec{ \tau_s }$, i.e., the Stokes parameters, are the coefficients of the expansion. Therefore, $- \imath \hat U^{\dagger} \hat U_{\omega}$ can be determined by experimentally measuring these Stokes parameters. Due to polarization mode dispersion a light signal in an arbitrary state $\ket{s}$ at the input of an optical fiber exhibits a mean signal time delay at the output given by the equation:
\cite{nelson2000measurement}:
\begin{eqnarray}
\tau_g = \tau_o + \vec{ \tau_s } \cdot \hat{ s },
\end{eqnarray}
where $\hat{s} = \braket{ s | \vec{\sigma} | s } = \{ s_1, s_2, s_3 \} $ is the Stokes vector of $\ket{s}$. Experimentally measuring the Stokes parameters can be accomplished by measuring mean signal time delays at the optical fiber output for given input states $\hat{s}$. Let's closely examine Eq. 2, specifically the inner product $\vec{ \tau_s } \cdot \hat{ m_s }$. When $\hat{s}$ is aligned to one of the Stokes parameters of $\vec{\tau_s}$ the mean signal time delay measured at the optical fiber output is the value of that Stokes parameter. In general, each pair of eigenstates of each Pauli matrix corresponds to each of the Stokes parameters \cite{goldstein2003polarized}. Launching a light signal at the optical fiber input as one of the eigenstates of $\hat \sigma_1$, $\hat \sigma_2$, or $\hat \sigma_3$ which are horizontal/vertical, diagonal/anti-diagonal, or right/left circular polarization, i.e., $s_1=\pm1$, $s_2=\pm1$, $s_3=\pm1$, will result in a mean time delay at the output that is the value of $\tau_1$, $\tau_2$, or  $\tau_3$, respectively. As can be seen the $SU(2)$ representation of $- \imath U^{\dagger} \hat U_{\omega}$, via the Stokes parameters, organizes a relationship between launch conditions at the optical fiber input and measurable mean signal time delays at the optical fiber output from which the principal states of polarization can be determined.  

This idea can be extended to a multimode optical fiber with $N$ modes where $ -\imath \hat U^{\dagger} \hat U_{\omega}$ is an $N \times N $ Hermitian matrix. Again, by merely exploiting its dimensionality and Hermiticity the operator $- \imath U^{\dagger} \hat U_{\omega}$ can be mathematically represented by the group $SU(N)$ and therefore expanded by generators of the corresponding Lie algebra. The Lie algebra for $SU(N)$ is $[\hat \lambda_i, \hat \lambda_j] = \imath f_{ijk} \hat \lambda_k$ and has $N^2-1$ generators where $f_{ijk}$ are the structure constants \cite{kaku1993quantum, georgi1982lie};  $ -\imath \hat U^{\dagger} \hat U_{\omega}$ can be expanded in terms of the generators given by $ -\imath \hat U^{\dagger} \hat U_{\omega} = \sum_{n=1}^{N^2 -1} \tau_n \hat \lambda_n $ where  $\tau_n$ and $\hat \lambda_n$ are the expansion coefficients and generators, respectively. To illustrate how this works we use the example of a \textit{few mode} optical fiber with a step index circular core consisting of the $HE_{11}^o$, $HE_{11}^e$, $TM_{01}$, $TE_{01}$, $HE_{21}^o$, and $HE_{21}^e$ eigenmodes. In this case, $N=6$  and $ -\imath \hat U^{\dagger} \hat U_{\omega}$ is a $6 \times 6$ Hermitian matrix. By merely exploiting its dimensionality and Hermiticity the operator $- \imath \hat U^{\dagger} \hat U_{\omega}$ can be mathematically represented by the group $SU(6)$ To further simplify the problem we look for irreducible representations of $SU(6)$ by recognizing that the modes can be broken up into \textit{mode groups} where $\{HE_{11}^e$, $HE_{21}^o\}$ and \{$HE_{21}^o$, $HE_{21}^e$, $TM_{01}$, $TE_{01}$\} are referred to as the $LP_{01}$ and $LP_{11}$ mode groups, respectively. To a good approximation coupling between mode groups is much less than coupling between the modes within each group \cite{bai2012mode}. Mathematically this means $ -\imath \hat U^{\dagger} \hat U_{\omega}$ can be block diagonalized and is given by $ -\imath \hat U^{\dagger} \hat U_{\omega} = \hat{LP}_{01} \bigoplus \hat{LP}_{11} $. The first block diagonal matrix $\hat{LP}_{01}$ is a $2 \times 2$ Hermitian matrix corresponding to the N=2 modes of the $LP_{01}$ mode group and can be represented by the group $SU(2)$ as described above. The second black diagonal matrix $\hat{LP}_{11}$ is a $4 \times 4$ Hermitian matrix corresponding to the N=4 modes of the $LP_{11}$ mode group and can be represented by the group $SU(4)$. The basis vectors for $\hat{LP}_{11}$ in terms of the $HE_{21}^o$, $HE_{21}^e$, $TM_{01}$, and $TE_{01}$ eigenmodes are:
\begin{eqnarray}
\ket{TM_{01}}  = \left(\begin{array}{cccc}1 \\ 0 \\  0 \\ 0\end{array}\right), \hspace{1mm} \ket{TE_{01}}  = \left(\begin{array}{cccc}0 \\ 1 \\  0 \\ 0\end{array}\right), \hspace{1mm} 
\ket{HE_{21}^{o,e}}  = \left(\begin{array}{cccc}0 \\ 0 \\  1 \\ 0\end{array}\right), \left(\begin{array}{cccc}0 \\ 0 \\  0 \\ 1\end{array}\right). \nonumber \\
\end{eqnarray}
$\hat{LP}_{11}$ can be expanded by the 15 generators of the corresponding $SU(4)$ Lie Algebra given by $\hat{LP}_{11} = \sum_{n=1}^{15} \tau_n \hat \lambda_n $. The generators are chosen as analogs of the Pauli matrices of Eq. 1 referred to as generalized Gell-Mann matrices; where the first eight are given by $(\vec{ \lambda } = \{ \hat  \lambda _1, ... , \hat \lambda _{15} \})$ \cite{kaku1993quantum, georgi1982lie}:
\begin{eqnarray}
\hat \lambda_1 = \left(\begin{array}{cccc}0 & 1 & 0 & 0 \\ 1 & 0 & 0 & 0 \\ 0 & 0 & 0 & 0 \\ 0 & 0 & 0 & 0 \end{array}\right), \hspace{5mm}
\hat \lambda_2 = \left(\begin{array}{cccc}0 & -\imath & 0 & 0 \\ \imath & 0 & 0 & 0 \\ 0 & 0 & 0 & 0 \\ 0 & 0 & 0 & 0 \end{array}\right), \nonumber 
\end{eqnarray}
\begin{eqnarray}
\hat \lambda_3 = \left(\begin{array}{cccc}1 & 0 & 0 & 0 \\ 0 & -1 & 0 & 0 \\ 0 & 0 & 0 & 0 \\ 0 & 0 & 0 & 0 \end{array}\right), \hspace{5mm} 
\hat \lambda_4= \left(\begin{array}{cccc}0& 0 & 1 & 0 \\ 0 & 0 & 0 & 0 \\ 1 & 0 & 0 & 0 \\ 0 & 0 & 0 & 0\end{array}\right), \nonumber 
\end{eqnarray}
\begin{eqnarray}
\hat \lambda_5= \left(\begin{array}{cccc}0 & 0 & -\imath & 0 \\ 0 & 0 & 0 & 0\\ \imath & 0 & 0 & 0 \\ 0 & 0 & 0 & 0\end{array}\right), \hspace{5mm} 
\hat \lambda_6 = \left(\begin{array}{cccc}0 & 0 & 0 & 0 \\ 0 & 0 & 1& 0 \\ 0 & 1 & 0 & 0 \\ 0 & 0 & 0 & 0 \end{array}\right), \nonumber 
\end{eqnarray}
\begin{eqnarray}
\hat \lambda_7 = \left(\begin{array}{cccc}0 & 0 & 0 & 0 \\ 0 & 0&  -\imath & 0  \\ 0 & \imath & 0 & 0 \\ 0 & 0 & 0 &  0\end{array}\right), \hspace{5 mm} 
\hat \lambda_8 = \frac{1}{\sqrt{3}} \left(\begin{array}{cccc}1 & 0 & 0 & 0 \\ 0 & 1 & 0 & 0 \\ 0 & 0 & -2 & 0 \\ 0 & 0 & 0 & 0 \end{array}\right).
\end{eqnarray}
The benefit of the $SU(4)$ representation of $\hat{LP}_{11}$ is in its ability to connect the generalized Gell-Mann matrices of Eq.  4 to Stokes parameters for the $HE_{21}^o$, $HE_{21}^e$, $TM_{01}$, and $TE_{01}$ eigenmodes which have been referred to as \textit{higher-order} Stokes parameters \cite{PhysRevLett.107.053601,PhysRevLett.108.190401}. As $\ket{m_s}$ is an eigenstate of $\hat{LP}_{11}$ its eigenvalue equation can be expressed in terms of the expansion given by $(\tau_s \cdot \hat \lambda) \ket{m_s} = (\tau_o +\tau_g) \ket{m_s}$ where $\vec{\tau_s}$ is a higher-order Stokes vector given by $\vec{ \tau_s } = \tau \braket{ m_s | \vec{ \lambda } | m_s} = \{ \tau_1, ..., \tau_{15} \} $ referred to as the \textit{mode dispersion vector}; the components of $\vec{ \tau_s }$, i.e., the higher-order Stokes parameters, are the coefficients of the expansion. Therefore, $\hat{LP}_{11}$ can be determined by experimentally measuring these higher-order Stokes parameters. Following Eq. 2, experimentally measuring the higher-order Stokes parameters can be accomplished by measuring mean signal time delays at the optical fiber output for given input states $\hat{s}$ where $\hat{s} = \braket{ s | \vec{\lambda} | s } = \{ s_1, ... , s_{15} \}$ is the higher-order Stokes vector of $\ket{s}$. As compared to a single mode optic fiber, in this case, each pair of eigenstates of each generalized Gell-Mann matrix corresponds to each of the higher-order Stokes parameters. Let's closely examine the eigenstates of the generalized Gell-Mann matrices of Eq. 6 using the basis vectors of Eq. 4. Embedded within are the more conventional Gell-Mann matrices associated with the group $SU(3)$. In fact, within $SU(4)$ there are two $SU(3)$ subgroups. In turn, within each $SU(3)$ subgroup there are three $SU(2)$ subgroups the first being $\{ \lambda_1, \lambda_2, \lambda_3 \}$ whose eigenstates are linear combinations of $\ket{TM_{01}}$ and $\ket{TE_{01}}$. The eigenstates of $\hat\lambda_1$ are $\ket{TM_{01}}$ and $\ket{TE_{01}}$ themselves which are referred to as radial and azimuthal polarization, respectively \cite{Zhan}. The eigenstates of $\hat\lambda_2$ are $\ket{TM_{01}} + \ket{TE_{01}}$ and $\ket{TM_{01}} - \ket{TE_{01}}$ which are referred to as spiral polarization \cite{gori2001polarization}. The eigenstates of $\hat\lambda_3$ are $\ket{TM_{01}} + \imath \ket{TE_{01}}$ and $\ket{TM_{01}} - \imath \ket{TE_{01}}$ which are right/left circular polarized optical vortices, i.e. angular momentum eigenstates of light \cite{volyar1996vortex,volyar1996vortical}. 
The eigenstates for the remaining Gell-Mann matrices of Eq. 4 are summarized in Table 1. The eigenstates of $\lambda_5$ and $\lambda_7$ are are referred to as hybrid vector polarization \cite{doi:10.1117/12.841920, milione2011hybrid}. The eigenstates for $\lambda_4$, $\lambda_6$, and $\lambda_8$ are linear polarized $LP_{11}$ modes of various linear polarization and mode rotation. The eigenstates of the remaining 7 generalized Gell-Mann matrices not shown in Eq. 5 can be found in a similar way. Launching a light signal at the optical fiber input as one of the eigenstates of $\hat \lambda_1, ..., \hat \lambda_{15}$, i.e., $s_1=\pm1$, ..., $s_{15}=\pm1$, will result in a mean time delay at the output that when measured is the value of $\tau_1, ... , \tau_{15}$, respectively. As can be see the $SU(4)$ representation of $\hat{LP}_{11}$, via the higher-order Stokes parameters, organizes a relationship between launch conditions at the optical fiber input and measurable mean signal time delays at the optical fiber output from which  principal modes can be determined. A similar pairwise measurement method for $N \times N$ systems of higher-order modes has recently been demonstrated \cite{giovannini2013characterization}.

Knowing $\ket{m_s}$, $\ket{m_s}$ is determined by $\ket{m_t} = \exp(\imath \phi_o )\hat U \ket{m_s}$. In a single mode optical fiber $\hat U$ or its Mueller matrix representation $\hat M$ can be determined using what is referred to as a Jones matrix eigenanalysis or a Mueller matrix method where the state of polarization of the light signal at the optical fiber output, $\vec{t}$, is measured via Stokes polarimetry as a function of input state of polarization $\vec{s}$ \cite{Jones:47, heffner1992automated, jopson1999measurement}. This idea can be extended to the $LP_{11}$ mode group described above where in this case $\hat{s} = \braket{ s | \vec{\lambda} | s }$ and $\hat{t} = \braket{ t | \vec{\lambda} | t }$ are input and output higher-order Stokes vectors, respectively. $\hat M$ can be determined by launching each higher-order Stokes parameter of $\hat{s}$ in Table 1 one at a time  with unit amplitude and then measuring $\hat{t}$ via the equivalent of Stokes polarimetry for higher-order modes. Each matrix element of $\hat M$ is given by $M_{ij} = t_{i,j} = $ where $i$ and $j$ correspond to the launched, $s_i$, and measured, $t_j$, Stokes parameters, respectively. Stokes polarimetry for higher-order modes is feasible and has been demonstrated \cite{flamm2013all, flamm2012mode, shapira2005complete}.
\begin{table}[htp]
\caption{Launch Conditions}
\begin{ruledtabular}
\begin{tabular}{ccc}
Generator & Stokes Parameter & Eigenstate  \\
\hline
$\hat \lambda_1$ & $s_1=\pm 1$ & $\ket{TM_{01}}$, $\ket{TE_{01}}$  \\
$\hat \lambda_2$ & $s_2=\pm 1$ & $\ket{TM_{01}} \pm \ket{TE_{01}}$  \\
$\hat \lambda_3$ & $s_3=\pm 1$ & $\ket{TM_{01}} \pm \imath \ket{TE_{01}}$ \\
$\hat \lambda_5$ & $s_7=\pm 1$ & $\ket{TM_{01}} \pm \imath \ket{HE_{21}^o}$  \\
$\hat \lambda_7$ & $s_8=\pm 1$ & $\ket{TE_{01}} \pm \imath \ket{HE_{21}^o}$ \\
$\hat \lambda_4$ & $s_4=\pm 1$ & $\ket{TM_{01}} \pm \ket{HE_{21}^o}$   \\
$\hat \lambda_6$ & $s_5=\pm 1$ & $\ket{TE_{01}} \pm \ket{HE_{21}^o}$  \\
$\hat \lambda_8$ & $s_6=\pm 1$ & $ \pm \ket{TM_{01}} \pm \ket{TE_{01}} \mp 2 \ket{HE_{21}^o}$  
\end{tabular}
\end{ruledtabular}
\end{table}
Experimentally generating the eigenstates of Table 1 is feasible using, for example, liquid crystal on silicon spatial light modulators \cite{flamm2013all, tripathi2012versatile}, $0-\pi$-phase plates \cite{bai2012mode}, liquid crystal q-plates  \cite{slussarenko2011tunable,cardano2012polarization}, or several all optical fiber methods \cite{milione2010stokes, milione2011hybrid, Milione:11, doi:10.1117/12.872004, doi:10.1117/12.807869}. In particular, tunable liquid crystal q-plates can efficiently generate arbitrary linear combinations of $TM_{01}$ and $TE_{01}$ modes over a broad range of frequencies. As principal modes depend directly on mode dispersion which changes with time an \textit{adaptive} technique for their determination is required and is feasible \cite{shen2005compensation}. 
 
Ideally, this method can be extended to multimode optical fibers with an arbitrary number of modes and $SU(N)$ representations. $ -\imath \hat U^{\dagger} \hat U_{\omega}$ can be continually block diagonalized into irreducible representations of mode groups \cite{6227317}, e.g., $ -\imath \hat U^{\dagger} \hat U_{\omega} = \hat{LP}_{01} \bigoplus \hat{LP}_{11} \bigoplus \hat{LP}_{12} \bigoplus \hat{LP}_{02} \bigoplus ... \bigoplus \hat{LP}_{\ell m}$. There are $N^2-1$ measurements re This method can also be extended to specialty multimode optical fibers, for example, an elliptical core optical fiber with $N=6$ modes. In this case, the purposeful elliptical deformation of the core imposes birefringence onto both the $LP_{01}$ and $LP_{11}$ mode groups \cite{snyder1983optical}. The irreducible representations of $- \imath U^{\dagger} \hat U_{\omega}$ correspond to the $LP_{01}$, $LP_{11}^o$, and $LP_{01}^e$ mode groups each being $2 \times 2$ Hermitian matrices represented by $SU(2)$. Other examples include spun elliptical core optical fibers where the mode groups correspond to angular momentum of light eigenstates \cite{alexeyev2012optical}, and MDM in silicon photonics \cite{driscoll2013asymmetric}.
 
Principal states of polarization in single mode optical fibers are associated with decoherence free subspaces for single photons entangled in polarization \cite{PhysRevLett.106.080404}. Principal modes may be associated with decoherence free subspaces in a higher dimensional Hilbert space for higher-order eigenmodes \cite{giovannini2013characterization}. In comparison, other methods to circumvent mode dispersion which correct mode dispersion after the data is received, such as a technique borrowed from radio communication referred to as MIMO, may be incompatible with entanglement based quantum information protocols \cite{bai2012mode}.
  
The authors acknowledge financial support from NSF GRFP Grant No. 40017-00-04, Corning Inc., and ARO Grant No. W911NF-09-1-0552.

\bibliographystyle{apsrev4-1}
\bibliography{apssamp}

\end{document}